\documentclass[prl,preprint,12pt,onecolumn,tightenlines]{revtex4}
\input{tcilatex}

\begin{document}

\title{A conditional-phase switch at the single-photon level}
\author{Kevin J. Resch, Jeffrey S. Lundeen, and Aephraim M. Steinberg}
\address{Department of Physics, University of Toronto\\
60 St. George Street, Toronto ON\ M5S 1A7\\
CANADA}

\begin{abstract}
We present an experimental realization of a two-photon conditional-phase
switch, related to the ``$c$-$\phi $'' gate of quantum computation. This
gate relies on quantum interference between photon pairs, generating
entanglement between two optical modes through the process of spontaneous
parametric down-conversion (SPDC). \ The interference effect serves to
enhance the effective nonlinearity by many orders of magnitude, so it is
significant at the quantum (single-photon) level. \ By adjusting the
relative optical phase between the classical pump for SPDC and the pair of
input modes, one can impress a large phase shift on one beam which depends
on the presence or absence of a single photon in a control mode.
\end{abstract}

\maketitle

A great deal of effort has gone into the search for a practical architecture
for quantum computation. As was recognized early on, single-photon optics
provides a nearly ideal arena for many quantum-information applications \cite%
{photon1}; unfortunately, the absence of significant nonlinear effects at
the quantum level (photon-photon interactions) appeared to limit the
usefulness of quantum optics to applications in communications as opposed to
computation. (Nevertheless, two recent proposals \cite{linear,linear2} have
resurrected the possibility of quantum computation using purely linear
optics.) Therefore, work has focused on NMR \cite{NMR1}, solid-state \cite%
{solid state}, and atomic \cite{kimble,QED1,ion trap1,atomic1} proposals for
quantum logic gates, but so far none of these systems has demonstrated all
of the desired features such as strong coherent interactions, low
decoherence, and straightforward scalability. Typical optical nonlinearities
are so small that the dimensionless efficiency of photon-photon interactions
rarely exceeds the order of 10$^{-10}$. We have recently been studying
techniques which use quantum interference to enhance these nonlinearities by
as much as ten orders of magnitude, leading to near-unit-efficiency
sum-frequency generation of individual photon pairs \cite{us switch}. In
this paper, we demonstrate that a similar experimental geometry can be used
to make a conditional-phase switch. Our switch is very similar to an
enhanced Kerr or cross-phase-modulation effect, in which the presence or
absence of a single photon in one mode may lead to a significant phase shift
of the other mode. This is also similar to experiments performed in cavity
QED \cite{kimble} (and to theoretical proposals for atomic vapours, in
systems relying on atomic coherence effects \cite{atomic coherence} or
photon-exchange interactions \cite{photon-exchange}), but occurs in a
relatively simple and robust system relying only on beams interacting in a
nonresonant nonlinear crystal.

The controlled-phase or $c$-$\phi $ gate performs the mapping 
\begin{eqnarray}
\left| 0\right\rangle _{1}\left| 0\right\rangle _{2} &\longrightarrow
&\left| 0\right\rangle _{1}\left| 0\right\rangle _{2}  \nonumber \\
\left| 0\right\rangle _{1}\left| 1\right\rangle _{2} &\longrightarrow
&\left| 0\right\rangle _{1}\left| 1\right\rangle _{2}  \nonumber \\
\left| 1\right\rangle _{1}\left| 0\right\rangle _{2} &\longrightarrow
&\left| 1\right\rangle _{1}\left| 0\right\rangle _{2}  \nonumber \\
\left| 1\right\rangle _{1}\left| 1\right\rangle _{2} &\longrightarrow
&e^{i\phi }\left| 1\right\rangle _{1}\left| 1\right\rangle _{2},
\label{cphi}
\end{eqnarray}%
where the subscripts ``1'' and ``2'' indicate the two qubits, stored in two
distinct optical modes, and $\left| 0\right\rangle $ and $\left|
1\right\rangle $ represent zero- and one-photon states, respectively \cite%
{nielsenchuang}. \ Although in quantum mechanics an overall phase factor is
meaningless, this unitary transformation is non-trivial when we consider
what happens to superpositions of photon number. \ The operation will induce
a relative phase of $\phi $ \emph{between} the $\left| 0\right\rangle $ and $%
\left| 1\right\rangle $ states of qubit 2, if and only if qubit 1 is in
state $\left| 1\right\rangle $. (It is this \emph{relative} phase which is
referred to as the ``optical phase'' of mode 2 \cite{optphase1}.) \ 

Since our experiment relies on interference, its operation is sensitive to
the phase and amplitude of the initial state, and we must limit ourselves to
a specific set of inputs. In particular, we illuminate our switch with two
classical fields in weak coherent states,%
\begin{eqnarray}
\left| \Psi \right\rangle &=&\left| \alpha \right\rangle \otimes \left|
\beta \right\rangle \\
&\approx &\left[ \left| 0\right\rangle _{1}+\alpha \left| 1\right\rangle _{1}%
\right] \otimes \left[ \left| 0\right\rangle _{2}+\beta \left|
1\right\rangle _{2}\right] ,  \label{initial state}
\end{eqnarray}%
for $\left| \alpha \right| $, $\left| \beta \right| \ll 1$. This state
includes contributions of all four two-qubit computational-basis states from
Eq. \ref{cphi}. As we will show theoretically and experimentally, the
lowest-order action of the gate is to shift the phase of only the $\left|
1\right\rangle _{1}\left| 1\right\rangle _{2}$ state, as desired for $c$-$%
\phi $ operation. We discuss several different regimes of operation, where
the value of $\phi $ may be smaller or greater, accompanied by certain
tradeoffs related to other characteristics of the gate.

This gate differs from the canonical $c$-$\phi $ concept in several regards.
\ Principally, the input cannot be in a pure Fock state (e.g., $\left|
1\right\rangle _{1}\left| 1\right\rangle _{2}$), or an arbitrary
superposition of the computational-basis states, because the appropriate
relative phase of $\left| 0\right\rangle _{1}\left| 0\right\rangle _{2}$ and 
$\left| 1\right\rangle _{1}\left| 1\right\rangle _{2}$ must be chosen at the
outset. Nevertheless, the gate produces significant entanglement at the
output, and may be useful in non-deterministic operation \cite{linear}; in
other words, it may be possible to post-select the desired value of a given
qubit rather than supplying it at the input. Alternatively, such a gate
might be used in the polarization rather than the photon-number basis. The
interaction can be controlled through phase-matching conditions such that
the phase shift is only impressed if both photons have, for example,
vertical polarization. Thus, two-photon entangled states as typically
produced in down-conversion systems, which are more properly described as%
\begin{equation}
\left| \Psi \right\rangle =\left| 0\right\rangle \left| 0\right\rangle
+\varepsilon \left\{ a\left| H\right\rangle \left| H\right\rangle +b\left|
H\right\rangle \left| V\right\rangle +c\left| V\right\rangle \left|
H\right\rangle +d\left| V\right\rangle \left| V\right\rangle \right\}
\label{general}
\end{equation}%
could store the amplitudes of the four computational-basis states in the
amplitudes $a$, $b$, $c$, and $d$, with the (small) coefficient $\varepsilon 
$ ensuring that $\varepsilon d$ exhibits the appropriate phase relationship
between the vacuum and the vertically-polarized photon pair. \ Although the
vacuum term would dominate, as in most down-conversion experiments, the
computation would have the desired effect contingent simply on the eventual
detection of a photon pair. Remaining weaknesses of the scheme include
contamination due to states outside the computational basis (e.g., states in
which two photons are present in the same mode), but by operating in the
low-photon-number regime, it is straightforward to reduce such effects to a
negligible level. Finally, the question as to whether the entanglement
produced by these interactions might be useful as a generalized quantum gate
in some larger Hilbert space (including, in particular, these states of
higher photon number) remains open.

Our experiment is easily understood as a modified Mach-Zehnder
interferometer (MZI) (Fig. 1). \ The input beam is a weak laser pulse
(containing much less than one photon per pulse on average) which enters the
interferometer and is split into the signal (mode 1) and phase reference
(mode 3). \ Modes 1 and 3 are recombined at a beamsplitter after mode 1
passes through a $\chi ^{(2)}$ nonlinear crystal which is simultaneously
illuminated by a pump beam at frequency 2$\omega $. \ The output fringes
from the MZI serve to measure the relative phase introduced between the two
arms by the action of the crystal. \ Our control beam (mode 2) is another
very weak coherent state that crosses mode 1 inside the nonlinear crystal. \
Modes 1, 2, and 3 all have the same frequency $\omega $. \ Modes 1 and 2 are
chosen such that the classical pump laser will produce degenerate
down-converted photon pairs into them upon interaction with the nonlinear
crystal. \ Photon-counting detectors monitor one output of the
interferometer and mode 2. \ In order to demonstrate the conditional-phase
operation of the device, we measure the phase of the fringes at det. 1 and
compare the cases in which the control detector (det. 2) does or does not
fire. \ This `conditional homodyne' measurement \cite{conditional coherence}
is similar to recent studies of `wave-particle correlations' in cavity QED %
\cite{orozco}.

A more detailed schematic of the experiment is shown in Fig. 2. \ The beam
from a Ti:Sapph oscillator (centre wavelength 810nm, rep rate 80MHz, and
pulse duration 50fs) is used to create the four beams used in the
experiment. \ The phase reference, signal, and control beams are created by
separating a small amount of the fundamental beam with beamsplitters (BS) 3
and 1 -- all beamsplitters are 90/10 (T/R). \ The signal and control beams
are made by rotating the polarization after BS1 and treating the horizontal
and vertical components independently. \ All three of these beams are
subsequently attenuated using ND\ filters. \ The majority of the pump
undergoes second-harmonic generation (SHG) in a type-I $\beta $-barium
borate (BBO) crystal. \ With the fundamental removed, this 405-nm pulse
serves as the bright (classical) pump laser for parametric down-conversion.
\ The signal and control beams are recombined with the pump laser at BS 4
and all three beams are focused onto a second 0.5-mm BBO\ crystal
phase-matched for type-II\ down-conversion and, therefore, type-II SHG. \
The spot created on the down-conversion crystal is imaged through a spatial
filter to select a single spatial mode and improve the spatial overlap
between down-converted and laser beams \cite{us switch}. \ The output from
the spatial filter is separated by a polarizing beam splitter (PBS)\ such
that the vertically-polarized photons (the control beam) are sent to
detector 2 for direct photodetection, while the horizontally-polarized
signal beam interferes with the phase reference (whose polarization is
rotated to horizontal) at BS\ 2. \ Detector 1 measures the output from one
port of BS 2. \ Both detectors are silicon avalanche photo-diodes operating
in the Geiger mode (Perkin-Elmer SPCM-AQR-13). \ Interference filters, with
centre wavelengths of 810nm and bandwidths of 10nm, are placed in front of
each detector to ensure good spectral overlap between the down-converted and
laser beams.

In previous work \cite{us switch}, we demonstrated that quantum interference
leads to a phase-sensitive photon-pair production rate in a similar
geometry. \ The interference can be understood as follows. \ Initially,
modes 1 and 2 contain weak coherent states and mode p contains an intense
(classical) pump laser. \ The initial state of modes 1 and 2 can be
described as shown in Eq. \ref{initial state}. \ Under the interaction
Hamiltonian, $\mathcal{H}_{int}=ga_{1}^{\dagger }a_{2}^{\dagger
}a_{p}+g^{\ast }a_{1}a_{2}a_{p}^{\dagger }$, the lowest order action of the
pump laser is simply to add an amplitude for a photon pair through
parametric down-conversion. \ The state from Eq.\ref{initial state} then
evolves to%
\begin{equation}
\left| \Psi \right\rangle =\left| 00\right\rangle +\alpha \left|
10\right\rangle +\beta \left| 01\right\rangle +\left( \alpha \beta
+A_{DC}\right) \left| 11\right\rangle ,  \label{state}
\end{equation}%
where $A_{DC}$ is the amplitude for down-conversion. \ This amplitude is
proportional to both the pump laser amplitude, $\gamma $, and the coupling
constant, $g$. \ In the previous experiment, we observed the modulation in
the photon pair production rate by performing direct photon coincidence
counting on modes 1 and 2. \ We changed the phase of the amplitude $A_{DC}$
by changing the delay of the pump laser and, in so doing, changed the value
of $\left| \alpha \beta +A_{DC}\right| ^{2}$ -- the probability of producing
a photon pair$.$ \ However, this process also affects the \emph{phase} of
that amplitude, i.e. $\arg \left( \alpha \beta +A_{DC}\right) .$ \ This is
the `cross-phase modulation' we study; in particular, for $\left|
A_{DC}\right| \ll \left| \alpha \beta \right| ,$ its principal effect is to
shift the phase of the $\left| 11\right\rangle $ term, as desired for a c-$%
\phi .$ \ Of course, the absolute phase of a state is never experimentally
observable; we therefore study the relative phase between $\left|
11\right\rangle $ and $\left| 01\right\rangle ,$ contrasting it with the
case of no control photon: $\left| 10\right\rangle $ vs. $\left|
00\right\rangle .$ \ This relative phase is precisely the optical phase
measured by our Mach-Zehnder interferometer. \ We rewrite the state in Eq. 5
in terms of modes 1 and 2 as follows: 
\begin{equation}
\left| \Psi \right\rangle =\left( \left| 0\right\rangle _{1}+\alpha \left|
1\right\rangle _{1}\right) \left| 0\right\rangle _{2}+\beta \left[ \left|
0\right\rangle _{1}+\left( \alpha +\frac{A_{DC}}{\beta }\right) \left|
1\right\rangle _{1}\right] \left| 1\right\rangle _{2}.  \label{final state 1}
\end{equation}%
In this form, it is evident that entanglement is generated between the
photon number in mode 2 and the optical phase in mode 1; the conditions that 
$\left| \alpha \right| ,$ $\left| \beta \right| \ll 1$ limit the state to
one of nonmaximal entanglement. \ However, this device can be used to
manipulate the photon pair amplitudes in the general down-conversion state
(Eq. \ref{general}) to generate maximal entanglement (i.e. a Bell state) in
the coincidence subspace \cite{bellstateus}. \ The relative magnitudes of
the different amplitudes for a photon pair determine different regimes of
operation. \ In the regime where $\left| A_{DC}\right| \ll \left| \alpha
\beta \right| ,$ (i.e. the probability of a down-conversion event is much
less than the ``accidental'' coincidence rate from the signal and control
beams) there is a small phase shift but, to first order, no change in rate.
\ In the opposite limit where $\left| A_{DC}\right| >\left| \alpha \beta
\right| ,$ the maximum phase-shift is $180^{\circ }$ and it occurs at the
point of maximum destructive interference. \ 

To explore the small phase-shift regime, we adjusted our signal and control
beam intensities to obtain, in the absence of interference, a coincidence
rate of $\left( 256\pm 3\right) $ s$^{-1}$ between Det. 1 and Det. 2. \ Our
coincidence rate from down-conversion alone was $\left( 4.7\pm 0.2\right) $ s%
$^{-1}$. \ The singles rates at Det. 1 (again in the absence of
interference) were 88$\times 10^{3}$ s$^{-1}$ from the signal beam alone and
79$\times 10^{3}$ s$^{-1}$ from the phase reference; Det. 2 received a
singles rate of 282$\times 10^{3}$ s$^{-1}$ from the control beam. \ This
corresponds to several photons per thousand laser pulses. \ The singles
rates due to down-conversion were 400 s$^{-1}$ at Det. 1 and 300 s$^{-1}$ at
Det. 2. \ To perform the experiment, the phase reference was blocked and
pump delay moved in sub-wavelength steps to observe fringes in the photon
pair production rate (described in \cite{us switch}). \ The delay was
stopped at a maximum in that rate, providing the definition of zero delay. \
With the pump delay fixed, we scanned over a few Mach-Zehnder interference
fringes by stepping the reference delay in 0.04-$\mu $m steps, and recorded
the singles rates at the two detectors and their coincidence rate. \ Due to
the low probability of having a photon in any given control pulse, the
interference fringes in Det. 1's singles rate are dominated by the case
where zero photons are present in the control mode; the coincidence rate
shows the phase-shifted fringes when a control photon is detected. \ 

A sample data set is shown in Fig. 3 for a pump delay of $-1.6$ fs (about 455%
$^{\circ }$). \ For clarity, \ the fringes shown are taken in the large
phase-shift regime, with $\left| A_{DC}\right| >\left| \alpha \beta \right| $%
. \ To achieve this regime, we reduced our coincidence rate from the signal
and control beams to $\left( 1.1\pm 0.1\right) $ s$^{-1}$ in the absence of
interference; our down-conversion coincidence rate was $\left( 5.2\pm
0.2\right) $ s$^{-1}$. \ Det. 1 received about 700 s$^{-1}$ singles rate
from the signal and 8600 s$^{-1}$ from the phase reference; Det. 2 had a
singles rate of 129$\times 10^{3}$ s$^{-1}$ from the control beam. \ The
coincidence counts have been averaged over 40-second intervals due to the
considerable shot noise. \ The fringes were fit to cosine curves where the
period of the coincidence fringes were constrained to equal that of the
singles fringes. \ The phase difference was then extracted modulo 360$%
^{\circ }$. \ 

Relative phases were measured in this way for many different pump phase
delays; those values are summarized in Fig. 4. \ The phase-shifts measured
for the low phase-shift regime are the open circles (right-hand scale). \
The dashed line is the theoretical prediction based on the experimentally
observed ratio of coincidence rates, with no adjustable parameters. \ In
this regime, the phase shift is limited to approximately $\left|
A_{DC}\right| /\left| \alpha \beta \right| $ -- about 8$^{\circ }$ for the
experimental ratio of coincidence rates. \ The phase shift is approximately
sinusoidal in the pump phase for this ratio. \ The shifts in the large
phase-shift regime are shown in Fig. 4 as solid circles (left-hand scale). \
Theory is shown as a solid line and, again, involves no free parameters. \
It is clear that in this regime we are able to access any phase shift
(modulo 360$^{\circ }$). \ In this regime, the phase shift does not follow a
sinusoidal modulation but rather increases monotonically with the pump
phase, modulo 360$^{\circ }$. \ In the extreme limit, where $\left|
A_{DC}\right| \gg \left| \alpha \beta \right| $, the phase shift is equal to
the pump phase delay. \ There is strong agreement between theory and
experiment, with slightly reduced phase shifts in the low phase-shift regime
possibly attributable to background.

We have demonstrated the correlation between the photon number in one mode
and the optical phase in another in a coherent conditional-phase switch. \
Our theoretical description of the device shows that entanglement between
the two modes is generated, but explicit demonstration requires additional
measurements. \ This is a new type of asymmetric entanglement \cite%
{conditional coherence}, of the sort required for the quantum $c$-$\phi $
gate. \ However our switch differs from the $c$-$\phi $, since the switch's
reliance on quantum interference makes it intrinsically dependent on the
optical phase of the input beams. \ While this phase-dependence will not
allow the gate to operate on Fock states, the gate does act exactly as a $c$-%
$\phi $ in the coincidence basis in some interesting situations; these
situations will be the subject of future work. \ Methods like the one
described in this paper of creating and controlling entanglement at the
single-photon level are very exciting for the field of quantum optics and
are promising steps towards all-optical quantum computing.

The authors thank Andrew White, Christina Pencarski, and Daniel Lidar for
thought-provoking discussion. \ K.R. acknowledges financial support of the
Walter C. Sumner Foundation. \ This work was funded by Photonics Research
Ontario and NSERC.

\bigskip

\textbf{Figure Captions}

\textbf{Fig. 1 -- A cartoon of the experiment. \ The signal beam, a weak }$%
\left( \left| \alpha \right| \ll 1\right) $ \textbf{coherent state, is
passed through a Mach-Zehnder interferometer in order to measure the phase
shift. \ This shift is imprinted by a }$\chi ^{(2)}$ \textbf{crystal pumped
with a strong classical pump (p), only when the control beam (also a weak
coherent state with mean photon number }$\left| \beta \right| ^{2}\ll 1$%
\textbf{) contains a photon. \ This conditional-phase operation is verified
by correlating the MZ output fringes at Det. 1 with detection of a control
photon at Det. 2.}

\textbf{Fig. 2 -- Schematic of the experiment. \ BS 1--4 are 90/10 (T/R)
beam splitters; SHG\ consists of two lenses and a 0.1-mm BBO\ crystal for
type-I second harmonic generation; }$\mathbf{\lambda /2}$ \textbf{are
half-wave plates; S.F. is a spatial filter; I.F. are interference filters
centred at 810nm with a 10nm bandwidth; BG is a coloured glass filter; PBS\
is a polarizing beam splitter; Det. 1 and 2 are single-photon counting
modules. \ The pump laser at 405nm is separated from the 810nm light by
using a fused-silica prism before the S.F. -- this is not shown for clarity.}

\textbf{Fig. 3 -- Phase shifted fringes in the large phase-shift regime. \
The Det. 1 singles rate (open squares, dashed line) and coincidence rate
between Det. 1 and Det. 2 (closed circles, solid line) are shown as a
function of the reference delay. \ The coincidence fringes display the phase
of the signal for cases in which a control photon was present; the singles
are dominated by cases in which no photon was present. \ The coincidence
counting rate\emph{\ lags} the singles rate by }$\left( \mathbf{65\pm 8}%
\right) ^{\mathbf{\circ }}$\textbf{. \ This data set was taken when the pump
phase lagged the phase of the LOs. \ }

\textbf{Fig. 4 -- Phase shift versus pump phase delay. \ The phase of the
pump laser was changed via the pump delay and was estimated using the
accompanying modulation in the mean coincidence rate \cite{us switch}. \ The
phase shift between the coincidence and singles fringes is plotted against
the pump phase delay for both the large phase-shift regime (solid circles)
and the small phase-shift regime (open circles). \ The solid and dashed
lines show the theoretical predictions for these two cases respectively,
based only on the measured ratio of the individual-path rates, and with no
adjustable parameters. \ }


\begin{thebibliography}{99}
\bibitem{photon1} C. H. Bennett and G. Brassard, \emph{Proceedings of the
IEEE International Conference on Computers, Systems \& Signal Processing,
Bangalore, India }(IEEE, New\ York, 1984), p. 175-179; A. K. Ekert, J. G.
Rarity, and P. R. Tapster, \emph{Phys. Rev. Lett.}, \textbf{69}, 1293
(1992); A. Muller, J. Breguet, and N. Gisin, \emph{Europhysics Letters}, 
\textbf{23}, 383 (1993); W. T. Buttler, et al., \emph{Phys. Rev. Lett.}, 
\textbf{81}, 3283 (1998); C. H. Bennett, et al., \emph{Phys. Rev. Lett.}, 
\textbf{70}, 1895 (1993); D. Bouwmeester, et al., \emph{Nature}, \textbf{390}%
, 575 (1997).

\bibitem{linear} E. Knill, R. Laflamme, and G. Milburn, \emph{Nature}, 
\textbf{409}, 46 (2001)

\bibitem{linear2} D. Gottesman, A. Kitaev, and J. Preskill, Phys. Rev. A,
64, 012310 (2001).

\bibitem{NMR1} N. A. Gershenfeld and I. A. Chuang, \emph{Science}, \textbf{%
275}, 350 (1997); J. A. Jones, M. Mosca, and R. H. Hansen, \emph{Nature}, 
\textbf{393}, 344 (1998); D. G. Cory, et al., \emph{Phys. Rev. Lett.}, 
\textbf{81}, 2152 (1998).

\bibitem{solid state} B. E. Kane, \emph{Nature}, \textbf{393}, 133 (1998).

\bibitem{kimble} Q. A. Turchette, C. J. Hood, W. Lange, H. Mabuchi, and H.
J. Kimble, \emph{Phys. Rev. Lett.}, \textbf{75}, 4710 (1995); A.
Rauschenbeutel, G. Nogues, S. Osnaghi, P. Bertet, M. Brune, J. M. Raimond,
and S. Haroche, \emph{Phys. Rev. Lett.}, \textbf{83}, 5166 (1999).

\bibitem{QED1} G. Nogues, et al., \emph{Nature}, \textbf{400}, 239 (1999);
P. W. H. Pinske, et al., \emph{Nature}, \textbf{404}, 365 (2000).

\bibitem{ion trap1} J. I. Cirac and P. Zoller, \emph{Phys. Rev. Lett.}, 
\textbf{74}, 4091 (1995); A. S\o rensen and K. M\o lmer, \emph{Phys. Rev.
Lett.}, \textbf{82}, 1971 (1999); C. Monroe, et al., \emph{Phys. Rev. Lett.}%
, \textbf{75}, 4714 (1995).

\bibitem{atomic1} G. K. Brennen, et al., \emph{Phys. Rev. Lett.}, \textbf{82}%
, 1060 (1999); D. Jaksch, et al., \emph{Phys. Rev. Lett.}, \textbf{85}, 2208
(2000).

\bibitem{us switch} K. J. Resch, J. S. Lundeen, and A. M. Steinberg, \emph{%
Phys. Rev. Lett.}, \textbf{87}, 123603 (2001), K. J. Resch, J. S. Lundeen,
and A. M. Steinberg, to be published in the \emph{Journal of Modern Optics}
(2001).

\bibitem{atomic coherence} S. E. Harris and L. V. Hau, \emph{Phys. Rev. Lett.%
}, \textbf{82}, 4611 (1999); M. M. Kash, et al., \emph{Phys. Rev. Lett.}, 
\textbf{82}, 5229 (1999).

\bibitem{photon-exchange} J. D. Franson, \emph{Phys. Rev. Lett.}, \textbf{78}%
, 3852 (1997).

\bibitem{nielsenchuang} M. A. Nielsen and I. L. Chuang, Quantum Computation
and Quantum Information\ (Cambridge University Press, Cambridge, 2000), p.
294.

\bibitem{optphase1} R. Loudon, The Quantum Theory of Light (2nd Edition),
(Clarendon Press, Oxford, 1973) p. 141-144; J. W. Noh, A. Foug\`{e}res, and
L. Mandel, \emph{Phys. Rev. Lett.}, \textbf{67}, 1426 (1991).

\bibitem{conditional coherence} K. J. Resch, J. S. Lundeen, and A. M.
Steinberg, \emph{Phys. Rev. Lett.}, \textbf{88}, 113601 (2002).

\bibitem{orozco} G. T. Foster, et al., \emph{Phys. Rev. Lett.}, \textbf{85},
3149 (2000).

\bibitem{bellstateus} K.J. Resch, J.S. Lundeen, and A.M. Steinberg,
quant-ph/0204034, to be published in Solvay conference proceedings (2002).
\end{thebibliography}
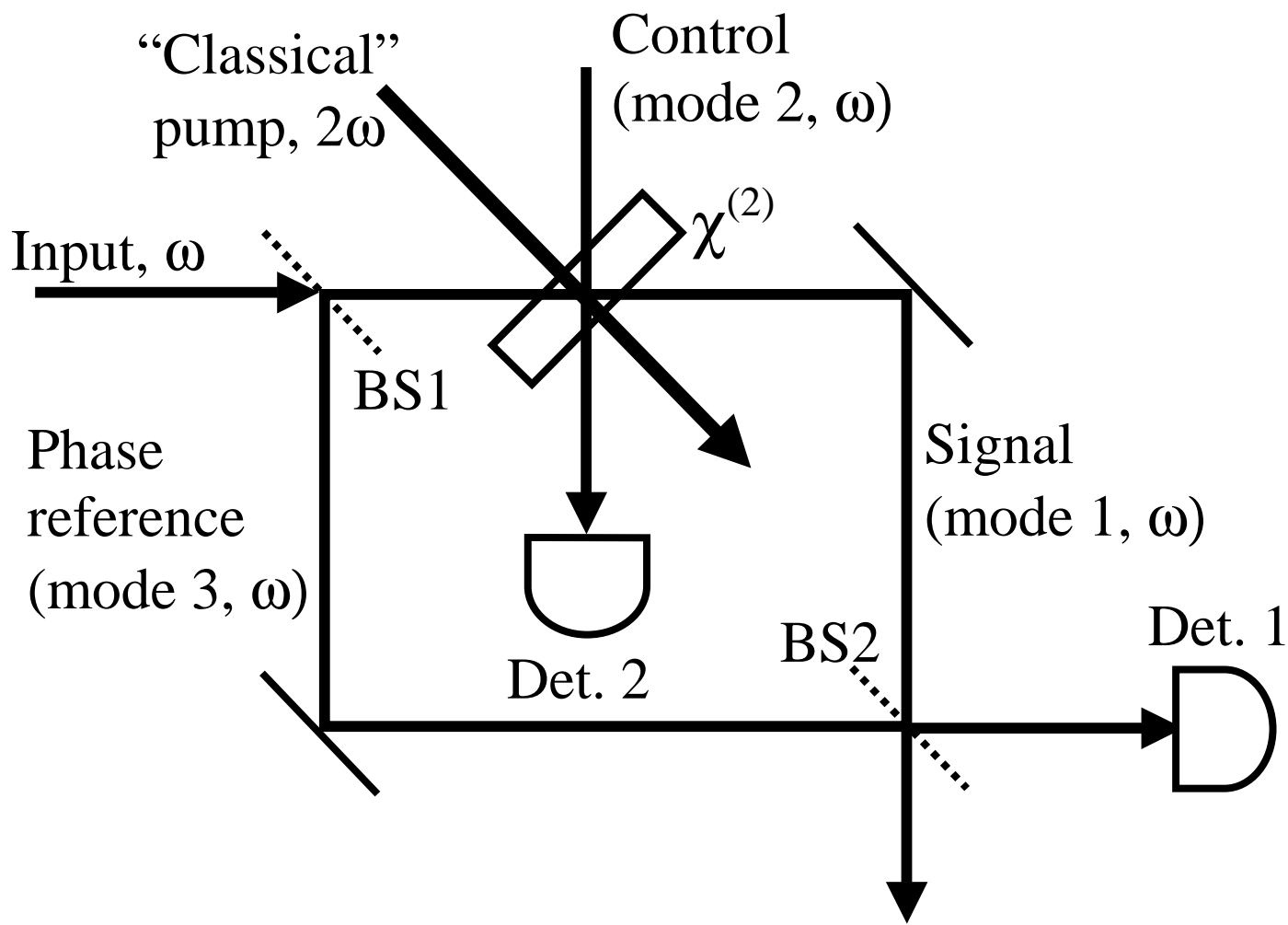
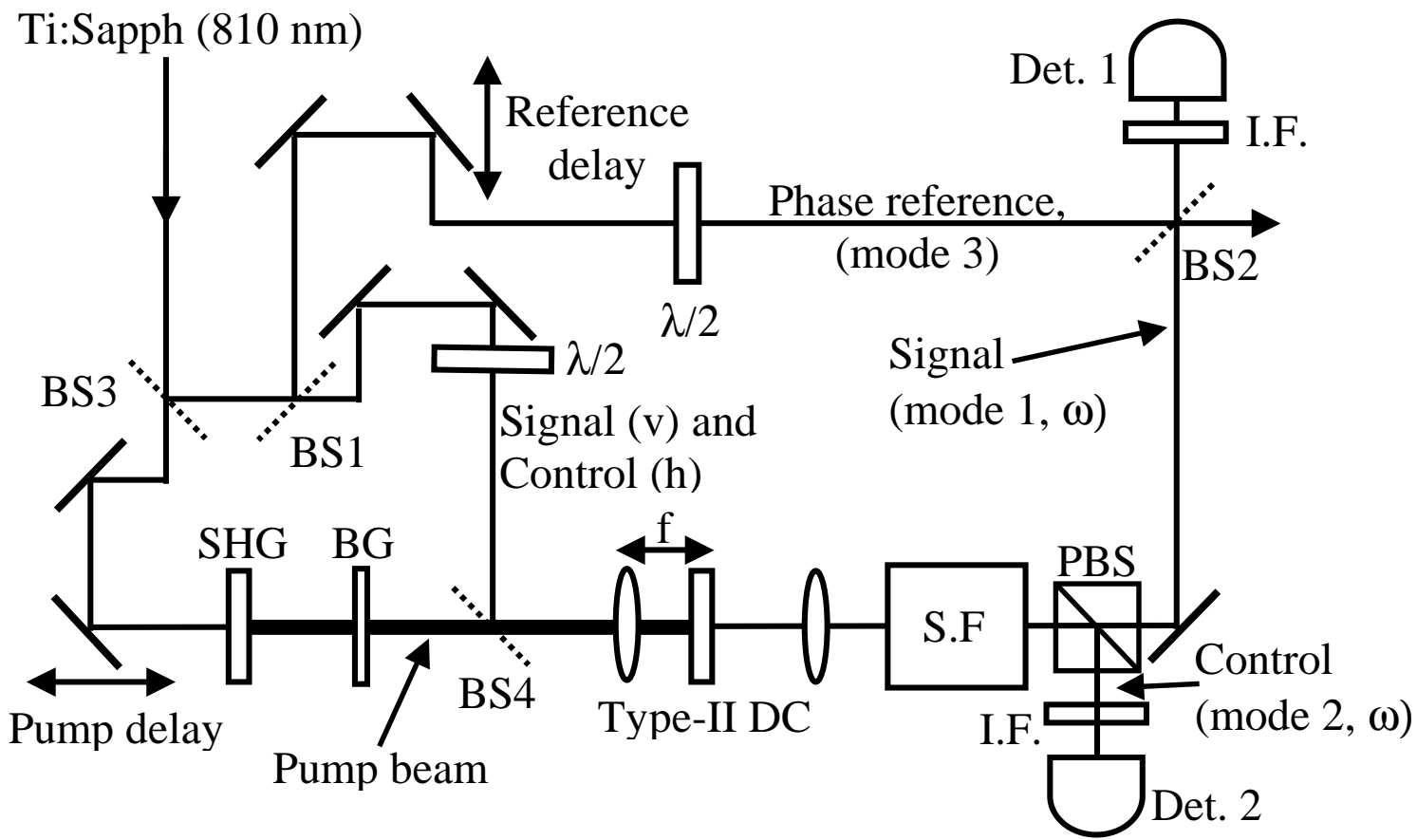
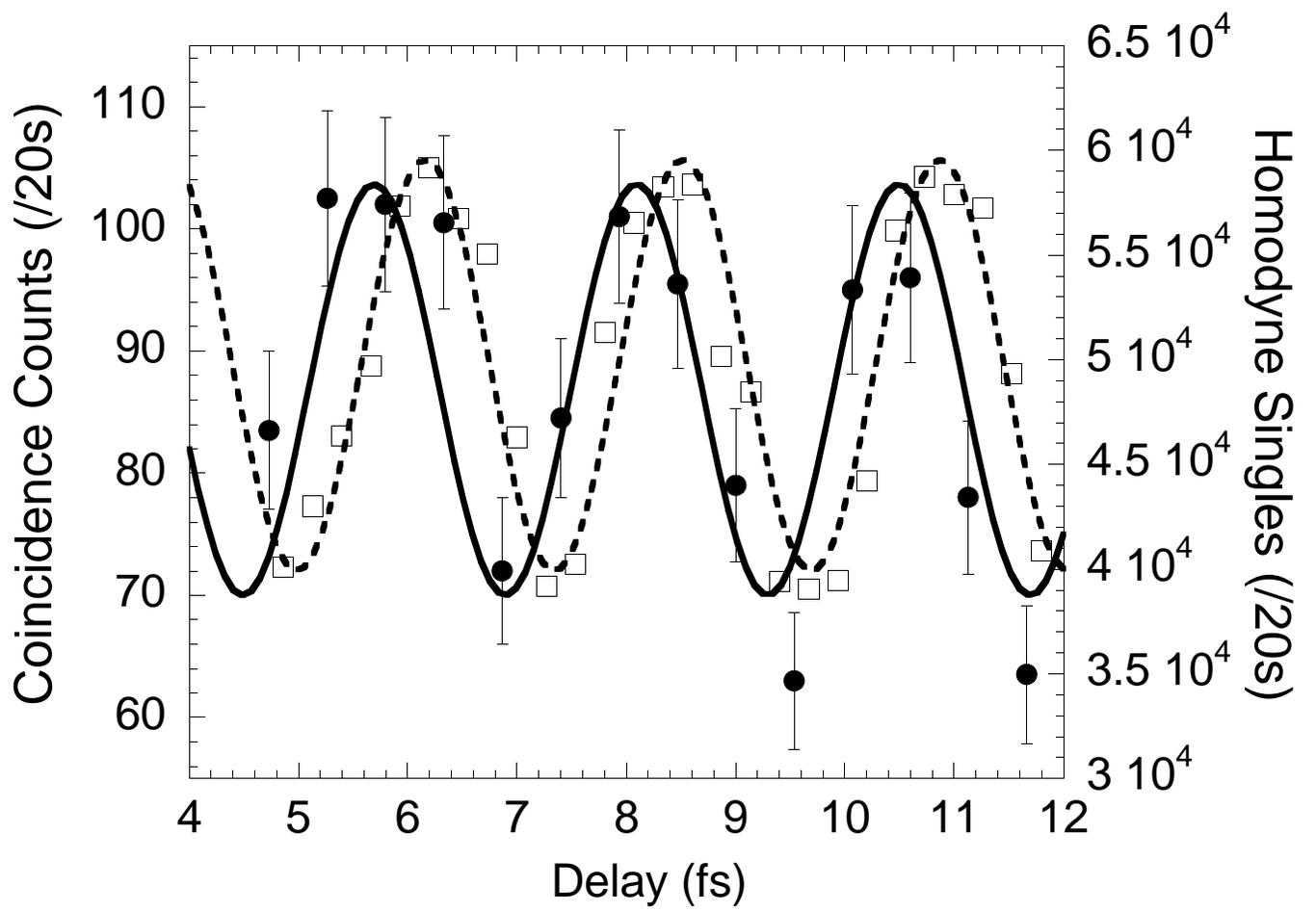
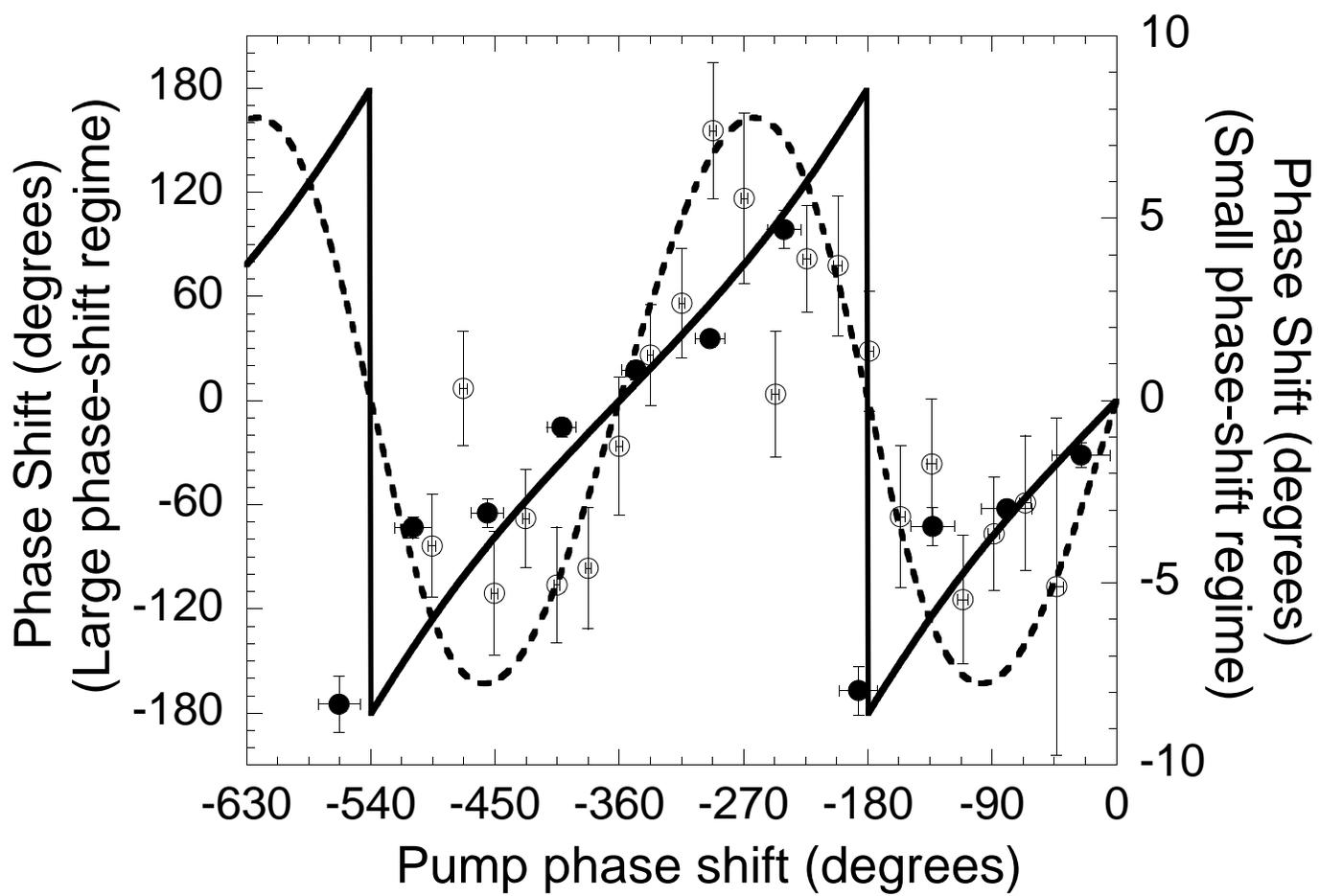
\end{document}